\documentclass[amsfonts,12pt]{article}

\usepackage{amsmath, amssymb, amsfonts, amsthm}
\date{}

\newcount\hour \newcount\minute
\hour=\time  \divide \hour by 60
\minute=\time
\loop \ifnum \minute > 59 \advance \minute by -60 \repeat
\def\nowtwelve{\ifnum \hour<13 \number\hour:
                      \ifnum \minute<10 0\fi
                      \number\minute
                      \ifnum \hour<12 \ A.M.\else \ P.M.\fi
         \else \advance \hour by -12 \number\hour:
                      \ifnum \minute<10 0\fi
                      \number\minute \ P.M.\fi}
\def\nowtwentyfour{\ifnum \hour<10 0\fi
                \number\hour:
                \ifnum \minute<10 0\fi
                \number\minute}

\title{Friedmann--Lemaitre Cosmologies \\via Roulettes and Other Analytic Methods}

\author{Shouxin Chen\footnote{Email address: chensx@henu.edu.cn}\\Institute of Contemporary Mathematics\\School of Mathematics\\Henan University\\
Kaifeng, Henan 475004, PR China\\\\
Gary W. Gibbons\footnote{Email address: gwg1@damtp.cam.ac.uk}\\
D. A. M. T. P.\\
University of Cambridge\\
Cambridge CB3 0WA, U. K.\\\\Yisong Yang\footnote{Email address: yisongyang@nyu.edu}\\Department of Mathematics\\Polytechnic School, New York University\\Brooklyn, New York 11201, U. S. A\\ \&\\NYU-ECNU
Institute of Mathematical Sciences\\New York University - Shanghai\\3663 North Zhongshan Road, Shanghai 200062, PR China}

\newcommand{\bfR}{{\Bbb R}}\newcommand{\arctanh}{{\mbox{arctanh}}}

\def\ben{\begin{equation}}
\def \een{\end{equation}}
\def \half {\frac{1}{2}}
\def \br {{\bf r}}
\def \bn {{\bf n}}

\def\XXint#1#2#3{{\setbox0=\hbox{$#1{#2#3}{\int}$}
 \vcenter{\hbox{$#2#3$}}\kern-.5\wd0}}

\newtheorem{oldtheorem}{Theorem}[section]
\newtheorem{oldassertion}[oldtheorem]{Assertion}
\newtheorem{oldproposition}[oldtheorem]{Proposition}
\newtheorem{oldremark}[oldtheorem]{Remark}
\newtheorem{oldlemma}[oldtheorem]{Lemma}
\newtheorem{olddefinition}[oldtheorem]{Definition}
\newtheorem{oldclaim}[oldtheorem]{Claim}
\newtheorem{oldcorollary}[oldtheorem]{Corollary}

\newcommand{\dd}{\mbox{\footnotesize    d}}
\newcommand{\ee}{\end{equation}}
\newcommand{\be}{\begin{equation}}\newcommand{\bea}{\begin{eqnarray}}
\newcommand{\eea}{\end{eqnarray}}
\newcommand{\e}{\mbox{e}}
\newcommand{\pa}{\partial}\newcommand{\Om}{\Omega}

\newcommand{\nn}{\nonumber}

\newcommand{\lm}{\lambda}
\def \cE{{\cal E}}
\begin{document}
\maketitle


\begin{abstract}
 In this work  a series of methods are developed for understanding the Friedmann equation when it is beyond the reach of the Chebyshev theorem.
First it will be demonstrated that every solution of the Friedmann equation admits a representation as a roulette such that information on the latter may be used to obtain that for the former. Next the Friedmann equation is integrated for a quadratic equation of state and for the Randall--Sundrum II 
universe, leading to a harvest of a rich
collection of new interesting phenomena. Finally an analytic method is used to isolate the asymptotic behavior of the solutions of the Friedmann equation, when the
equation of state is of an extended form which renders the integration impossible, and to establish
a universal exponential growth law.


\medskip


\end{abstract}

\maketitle

\tableofcontents

\section{Introduction}

When a homogeneous and isotropic perfect-fluid universe hypothesis is taken, the
Einstein equations of general relativity are reduced into a set of linear ordinary differential equations, governing the Hubble
parameter in terms of a scale factor, known as the Friedmann equations, which are of basic
importance in evolutionary cosmology. In the recent systematic works \cite{CGLY,CGY}, we explored a link between the Chebyshev theorem \cite{C0,C1} and the integrability of the Friedmann equations and obtained a
wealth of new explicit solutions when the barotropic equation of state is either linear or nonlinear,
in both cosmic and conformal times, and for flat and non-flat spatial sections
with and without a cosmological
constant. The purpose of the present paper  is to present
several analytic methods which may be used to
obtain either exact expressions or insightful knowledge of the solutions of the Friedmann equations
beyond the reach of the Chebyshev theorem. These methods may be divided into three categories:
roulettes, explicit integrations, and analytic approximations.

The idea of roulettes offers opportunities for obtaining some special solutions. Recall that many textbooks mention the fact that
the graph of the scale factor $a(t)$ against time $t$
for a  closed   Friedmann--Lemaitre
cosmology supported by a fluid whose pressure is negligible
follows is  cycloid,  thus providing  an attractive
image of the wheel of time steadily rolling from what
Fred Hoyle dubbed a Big Bang to a  Big Crunch reminding one
of Huyghens'  isochronous pendulum
and inviting speculations, such as those of Richard Chace Tolman
about cosmologies which are cyclic or cyclic up to a rescaling.
More recently  such models have been called Ekpyrotic
following an old tradition of the Stoics.

Few of those  textbooks give details or a derivation
of the roulette construction of a cycloid
and none, as far as we are  aware,
explain whether or not it is a mathematical fluke
or whether any plot of the scale factor against time
may be represented as a {\it roulette},
that is as the locus of a point on or inside
a curve which rolls without slipping along a straight line.

Evidence that this may be so is provided by
what is often called a {\it Tolman  Universe},
that is one in which the fluid is a radiation gas
whose energy density $\rho$ and Pressure $P$ are related
by $\rho=\frac{1}{3}P$. In this case, for a closed universes,
$a(t)$ is a semi-circle which of course may be obtained
as a sort of limiting case by  ``rolling without slipping''
the straight line interval given by it radius. Further evidence
is provided by De-Sitter spacetime which contains just
{\it Dark Energy}. In its closed form, the scale factor
has the form of a {\it catenary} which is the locus
of the focus of a {\it parabola} rolled without slipping
along its {\it directrix}.   In the light of this {\it lacuna}
in the literature, it is here proposed  to provide  a  general discussion,
showing among other things that a roulette construction is always
possible and how in principle to obtain the form of the rolling curve.
In order to do so  the article commences  by recalling some of the
classical theory of
plane curves, mainly developed in the $17 ^{\rm th}$ and $18 ^{\rm th}$
century, although its roots go back to the Greeks.  As the
theory is then applied to the {\it Friedmann and Raychaudhuri equations}
familiar in cosmology it will become apparent that
the ideas have a much wider relevance. In particular
they extend immediately to the motion
of light, or sound rays governed by {\it Snell's Law}
moving  in a vertical  stratified medium
with variable refractive index, and to the familiar central orbit
problems of mechanics.  More surprisingly perhaps
we may, as Delaunay first pointed out  in the $19^{\rm th}$ century,
we can also use the same description to describe the shapes of
axisymmetric soap films \cite{Delaunay,Eells}.
In this way we obtain a framework for
understanding in a unified  geometric fashion a wide variety of
iconic problems in physics.

Recall also that the Chebyshev theorem applies only to  integrals of binomial differentials \cite{CGLY,CGY}. In cosmology, one frequently encounters models which cannot be converted into
such integrals. Hence it will be useful
to present some methods which may be used
to integrate the Friedmann equations whose integrations
involve non-binomial differentials.
Here we illustrate our methods by integrating the model when the equation of state is quadratic
\cite{AB1,AB2,SSC}
and the Randall--Sundrum II universe \cite{Ge,RS,SWC,CC} with a non-vanishing cosmological constant for which both the energy density and its quadratic power are contributing to the right-hand side of the Friedmann equation. A quadratic equation of state introduces a larger degree of freedom for the choice of
parameters realized as the coefficients of the quadratic function. In
the flat-space case we show that for the solutions
of cosmological interest to exist so that the scale factor evolves from zero to infinity the coefficients
must satisfy a {\em necessary and sufficient condition} that confines the ranges of the parameters. We also
show that when  the quadratic term is present in
the equation of state the {\em scale factor cannot vanish at finite time}. In other
words, in this situation, it requires an infinite past duration
for the scale factor to vanish.
As another example that cannot be dealt with by the Chebyshev theorem, we study the Randall--Sundrum
II universe when the space is flat and $n$-dimensional and the cosmological constant $\Lambda$ is arbitrary. There are two cases of interest: (i) the equation of state is linear, $P=A\rho$, $A>-1$,and
(ii) the equation of state is that of the Chaplygin fluid type, $P=A\rho-\frac B\rho$, $A>-1, B>0$.
In the case (i) the big-bang solutions for $\Lambda=0$ and $\Lambda>0$ are similar as in the classical situation so that the scale factor either enjoys a power-function growth law or an exponential growth law according to whether $\Lambda=0$ or $\Lambda>0$. When $\Lambda<0$, however, we unveil the
fact that the solution has only a {\em finite lifespan} when $A\geq -1+\frac1n$, and we determine the
lifespan explicitly. When $-1<A<-1+\frac1n$, the solution is periodic, as in the classical case.
In the case (ii) with $\Lambda=0$, we find the explicit big-bang solution by
integration and display its
exponential growth law in terms of various physical parameters in the model. Finally, we carry
out a study of the big-bang solution for the Friedmann equation when the equation of state is of
an extended Chaplygin fluid form \cite{PK1,PK2,KKPMP}, $P=f(\rho)-\sum_{k=1}^m\frac{B_k}{\rho^{\alpha_k}}$, where
$f(\rho)$ is an analytic function and $B_1,\dots,B_m>0,\alpha_1,\dots,\alpha_m\geq0$, for which
an integration by whatsoever means would be impossible in general. We shall identify an
explicit range of $\Lambda$ for which the scale factor grows exponentially fast and deduce a universal formula for
the associated exponential growth rate. This formula covers all the explicitly known concrete cases.

The rest of this article is organized as follows. In \S 2 we review the Friedmann equation. In
\S 3 we demonstrate a relation between roulettes and the Friedmann equation and show how to use
this relation to find some special solutions. In \S 4 we consider the
integration of the Friedmann equation when the equation of state is quadratic. In \S 5 we solve
the Friedmann equation for the Randall--Sundrum II universe. In \S 6 we establish a universal
exponential growth law for the extended Chaplygin fluid universe. In \S 7 we summarize our results.
In Appendix (\S 8) we collect some relevant concepts and facts used in this article.

\section{Friedmann's equations}
\setcounter{equation}{0}

For generality we shall consider an $(n+1)$-dimensional  homogeneous and isotropic
spacetime with the line element
\be \label{1}
\dd s^2=g_{\mu\nu}\dd x^\mu\dd x^\nu=-\dd t^2+a^2(t)g_{ij} \dd x^i\dd x^j,\quad i,j=1,\dots,n,
\ee
where $g_{ij}$ is the  metric of an $n$-dimensional Riemannian manifold
$M$ of constant scalar curvature characterized by an indicator, $k=-1,0,1$,
so that $M$ is an $n$-hyperboloid, the flat space $\bfR^n$, or an $n$-sphere, with the
respective metric
\be \label{2}
g_{ij}\dd x^i\dd x^j=\frac1{1-kr^2}\,\dd r^2+r^2\,\dd\Om^2_{n-1},
\ee
where $r>0$ is the radial variable and $\dd\Om_{n-1}^2$ denotes the canonical metric of the unit
sphere $S^{n-1}$ in $\bfR^n$, and
 $t$ is referred to as the cosmological (or cosmic) time. The Einstein equations are
\be
G_{\mu\nu}+\Lambda g_{\mu\nu}=8\pi G_n T_{\mu\nu},
\ee
where $G_{\mu\nu}$ is the Einstein tensor, $G_n$ the universal gravitational constant in $n$ dimensions, and $\Lambda$ the cosmological constant, the speed of light is set to unity, and $T_{\mu\nu}$ is the
energy-momentum tensor. In the case of an ideal cosmological fluid, $T_{\mu\nu}$ is given by
\be \label{4}
T_{\mu}^{\nu}=\mbox{diag}\{-\rho_m,P_m,\dots,P_m\},
\ee
with $\rho_m$ and $p_m$ the $t$-dependent matter energy density and pressure.
Thus, assuming an ideal-fluid homogeneous and isotropic universe,
the Einstein equations are reduced into
the Friedmann equations
 \bea
H^2&=&\frac{16\pi G_n}{n(n-1)}\rho-\frac k{a^2},\label{5}\\
\dot{H}&=&-\frac{8\pi G_n}{n-1}(\rho+P)+\frac k{a^2},\label{6}
\eea
in which
$
H=\frac{\dot{a}}a
$
denotes the usual Hubble `constant' with $\dot{f}=\frac{\small{\dd} f}{\small{\dd} t}$, and $\rho,P$ are the effective energy
density and pressure related to $\rho_m,P_m$ through
\be \label{8}
\rho=\rho_m+\frac{\Lambda}{8\pi G_n},\quad P=P_m-\frac{\Lambda}{8\pi G_n}.
\ee
Besides, in view of (\ref{1}) and (\ref{4}) and (\ref{8}), we see that the energy-conservation law $\nabla_\nu T^{\mu\nu}=0$ reads
\be \label{9}
\dot{\rho}_m+n(\rho_m+P_m)H=0.
\ee

It is readily checked that (\ref{5}) and (\ref{9}) imply (\ref{6}). Hence
it suffices to consider (\ref{5}) and (\ref{9}).

In the rest of the paper, we omit the subscript $m$ in the energy density and pressure when there is
no risk of confusion.

\section{Friedmann's equations and roulettes}
\setcounter{equation}{0}

In \cite{CGY}, numerous Friedmann type equations arising in
a wide variety of applied areas of physics, other than cosmology, are integrated as well in view of the
Chebyshev theorem, which include the equation for light refraction  in a horizontally stratified medium, equations for soap films
and glaciated valleys, equation of catenary of equal strength, the elastica of Bernoulli and the capillary
curves, central orbit equations, equations of capillary curves, 
equations for spherically symmetric lenses, etc. These equations
may also be recast in the forms of roulettes as seen below.

\subsection{Rolling without slipping}

In what follows, extensive use has been made of
\cite{Oldrefs}. 
Suppose a closed convex curve $\gamma$
enclosing a domain $D$ with boundary
$\gamma = \pa D$   lies in a plane $\Pi$. Physically we think of $
D$ as a rigid {\it lamina} which rolls without slipping along the
a straight line $L$ in
another plane $\Pi^\prime$. We may suppose the line $L$ to be the $x$-axis
of a Cartesian coordinate system $(x,y)$ for the plane $\Pi ^\prime$.
The locus  in $\Pi^\prime$ of a fixed  point $O \in D$
defines a (single-valued) bounded (and hence no-monotonic)
periodic  graph    $\gamma ^\prime : y=y(x)$ over $L$ .
If $\gamma$ is smooth and  $O$ lies in the interior of $D$,
then $\tilde  \gamma$ will be a smooth simple curve. If $O \in \pa D $,
then $\gamma ^\prime$ will have cusps where $O$ meets  the $x$-axis.
If, by some contrivance, $O$ lies outside $D$, then the curve $\gamma ^\prime$
will no longer be a single-valued graph and it will have self-intersections.
If the point $O$ lies on the curve, one refers $\gamma'$ to a {\it cycloid}.
Otherwise to a {\it trochoid}.
Using $O$ as origin we may specify the  curve  $\gamma$
by its equation $\br=\br(t)$. We define $r=|\br|$  and
$p= \br \cdot \bn$ , where $\bn$ is the outward  normal of
of $\gamma$. Thus $r$ is distance from $O$ to the
point $\gamma(t)$ and $p$ the perpendicular distance from $O$ to the
tangent of the curve $\gamma$ at the point $\gamma(t)$.
The relation $(p\,,r) \in \mathbb{R} ^2 $ is called the {\it pedal equation}
of $\gamma$ with respect to the origin $O$. We also have a relation
$(y\,, y ^\prime) \in \mathbb{R}^2$, associated  to the
graph $\gamma$, where $y^\prime=\frac{\dd y}{\dd x}$.

Simple geometry gives
\ben
(p\,,r) = \left(y\,,y\sqrt{1+ {y^\prime}^2 }\right),\qquad (y,y ^\prime)= \left(p, \sqrt{
\frac{r^2}{p^2} -1 }\, \right). \label{pedal}
\een
If  $(r,\theta)$  are plane polar coordinates for the plane $\Pi $
with origin $O$, then
\ben
p= \frac{r}{\sqrt{1+ \frac{1}{r^2} \left( \frac{\dd
r}{\dd \theta}\right)^2 }} \,,\qquad
\Longleftrightarrow \qquad
\left(\frac{\dd u}{\dd \theta}\right)^2 + u^2 = \frac{1}{p^2}.
\een
Thus given a pedal curve we get a
relation $(y,y^\prime)$ and if we are fortunate we may be able
to solve for $y(x)$. Conversely given the graph $y(x)$
we may determine  the relation $(p,r)$ and if we are fortunate
we may be able to solve for $p=f(r)$ and hence  determine
the  equation of the pedal curve in in polar coordinates by quadratures.
There remains the problem of determining whether
the pedal curve is among the known plane curves
catalogued in various books, and if so what are its properties.

We conclude this section by recalling Steiner's theorem, which states that
\begin{itemize}
\item the arc  length along the rolling curve equals the arc length along the line on which it rolls.
\end{itemize} The area under the roulette is twice the area swept out by the
rolling curve.
\subsection{The Friedmann equation}
It is
convenient  in what follows to use units in which $8\pi G_3=3$ and $c=1$,
and to rewrite the Friedmann equation as
\ben
1+ (\dot a )^2 = a^2 \rho (a) + (1-k) a^2. \label{Friedmann2}
\een
Now  we may rescale $a$ and $t$
in (\ref{Friedmann2}) by the same constant
factor $f$ and transform the equation  to the form
\ben
1+ (\dot a )^2 = f^2 a^2 \rho (fa)  + f^2 (1-k) a^2. \label{Friedmann3}
\een
It follows from (\ref{pedal})
that the pedal equation of the roulette we seek is given by
\ben
r^2 = p^4 f^2 \rho(f p)   +  f^2 (1-k) p^4,
\een
where $f$ may be chosen as we wish.
Setting $k=1$ and $f=1$ gives our basic eqation:
\ben \boxed{\qquad
r^2 = p^4\rho (p)\qquad
\Longleftrightarrow \qquad \rho(p) = \frac{r^2}{p^4} \label{Friedmann}
\qquad }\een
which may be read in two directions. Either we know $\rho(a)$,
and we deduce the pedal equation, or given, a  pedal equation,
we deduce $\rho(a)$. In  the sequel we shall proceed in  both directions.

Before doing so we recall that in terms of conformal time $\eta$,
defined by $\dd t=a \dd\eta$,  the Friedmann
equation may be written (if $k=1$) as
\ben
\left(\frac{\dd a}{\dd\eta}\right) ^2 + k a^2 = a^4 \rho(a), \label{nonlin}
\een
which is a first integral of a non-linear oscillator 
equation obtained by differentiating (\ref{nonlin}) with respect
to $\eta$,
which is equivalent to the familiar  Raychaudhuri equation.

It is readily seen that the integrable cases in view of the Chebyshev theorem include the model
\be\label{1.8}
\rho(a)=\alpha a^{-2}+\beta a^{\sigma},
\ee
where $\alpha,\beta,\sigma$ are arbitrary constants.

If we introduce  \cite{Coq}
 the inverse scale factor, or redshift factor by
\ben
b=\frac{1}{a},
\een
we find
\ben
\left( \frac{\dd b}{\dd\eta} \right) ^2       + k b^2 = \rho\left(\frac{1}{b} \right).
\een
\subsubsection{Steiner's Theorem}

With this preparation we note that Steiner's theorem
implies that if we are  considering the scale factor
as a function of cosmic time, $a=a(t)$, the arc-length along
the rolling curve equal  to cosmic time and the area $\int a(t)\dd t$
swept out is twice  the area swept out on the rolling curve.
The area under the graph of the scale factor plotted against
time has no obvious meaning but if instead
we consider  scale factor
as a function of conformal  time, $a=a(\eta)$,
which  we   represented   as a roulette,  then arc-length of the rolling curve
would be equal to cosmic time and the area under the graph
$\int a(\eta)\dd \eta$  swept out would be equal to cosmic time and as
a consequence this would be twice the area swept out by the
rolling curve.

\subsection{A single-component fluid  in a closed universe}

In the simplest barotropic case with $k=1$ and $\gamma \ne \frac{2}{3}$
we may choose the scaling factor $f$  so that
\ben
\rho = \frac{1}{a^{3\gamma}}.
\een
In fact  John Barrow has pointed out \cite{Barrow} that
in this case,
if one  defines
\ben
g=   a^{\frac{3\gamma-2 }{2}},
\een
then the   Friedmann  equation becomes
\ben
\left(\frac{\dd g}{\dd \eta} \right ) ^2 + \left( \frac{3 \gamma-2}{2}
 \right )^2 g^2 = \left( \frac{3 \gamma-2}{2}
 \right )^2,
\een
whence we obtain the Raychaudhuri equation
\ben
\frac{\dd ^2 g}{\dd \eta ^2 }   + \left( \frac{3 \gamma-2}{2}
 \right )^2 g = 0, \label{Raychaudhuri}
\een
and so a general solution for $a$ as a function of
conformal time $\eta$  is available and cosmic time $t$ may then
be obtained in principle  by quadratures.

Not surprisingly, in view of the previous subsection and the above discussion, the same reduction may be carried out in the context of the slightly more general model (\ref{1.8}) with $\sigma=-3\gamma$ and $\gamma\neq\frac23$, which leads
to the updated equation
\be
\left(\frac{\dd g}{\dd\eta}\right)^2+(k-\alpha)\left(\frac{3\gamma-2}2\right)^2 g^2=\beta \left(\frac{3\gamma-2}2\right)^2,
\ee
and may be solved as before.

However we wish to construct  the  solution as a roulette.
The pedal equation is seen to be
\ben
r^2 = p^{4-3\gamma},
\een
which is the pedal equation of the curve
\ben
r^{\frac{3 \gamma-2}{4-3\gamma}} =
\cos \left(\frac{3 \gamma-2}{4-3\gamma} \theta
\right).
\een
Note that the $\gamma=\frac{2}{3}$  is a critical case
separating solutions which oscillate in conformal time
and those that blow up exponentially. Another special case is $\gamma=\frac{4}{3}$, radiation. Both require special treatment.

Some special cases are as follows.

\begin{itemize}
\item For dust, $\gamma=1$, $P=0$, we have
\ben
r= \cos \theta\,,
\een
which is the polar equation of a circle of radius $\half$, where the origin is
on its circumference.
\item If $\gamma = \frac{10}{9}$, $P= \frac{1}{9} \rho$,  we have
\ben
r= \cos (2 \theta),
\een
which is the polar equation of the Lemniscate of Bernoulli whose
Cartesian equation is
\ben
(x^2 + y^2 ) ^2 = x^2 -y^2.
\een
\item  For radiation,  $\gamma=\frac{4}{3}$, $P= \frac{1}{3}$,
we have
\ben r=1, \een
which is
which is the polar equation of a circle of radius $1$ , where the origin is
at its centre. However the locus of such a rolling curve would be
a straight horizontal line. This is not consistent with the
second-order Raychaudhuri equation (\ref{Raychaudhuri})
whose solution is
\ben
a= \sin \eta,\qquad  t= 1-\cos \eta.
\een

\item
For stiff matter,  $\gamma= 2$ , $P=\rho$,   we have
\ben
r^2 \cos (2 \theta) =1,
\een
which is the rectangular hyperbola
\ben
x^2-y^2=1,
\een
with the origin at its centre $(x,y)=(0,0)$.
\item If  $\gamma= \frac{8}{9}$, $P=-\frac{1}{9}\rho$,
we have
\ben
r= \half (1+\cos \theta),
\een
which is the polar equation of a cardioid
with origin at the cusp $\theta=\pi$.
\item If  $\gamma= \frac{5}{6}$, $P=-\frac{1}{6} \rho$,  we have
\ben
r= \cos ^3 \left(\frac{\theta}{3}\right),
\een
which is the polar equation of Cayley's sextic whose Cartesian equation is
\ben
 \left(  4(x^2+y^2) -x \right )  ^3 = 27 \left( x^2 + y^2 \right ) ^2.
\een
\item If  $\gamma= \frac{1}{3}$, $P=-\frac{2}{3}\rho$, 
 we have
\ben
r= \cos ^{-3} \left(\frac{\theta}{3}\right),
\een
which is the polar equation of
Tschirhausen's cubic,  whose  Cartesian equation
is
\ben
27 y^2 = (1-x)(x+8)^2.
\een
Tschirhausen's cubic is also known as
L'H\^opital's cubic or Catalan's trisectrix.
  \item
For a cosmological term, $\gamma=0$, 
\ben
\frac{2}{r} = 1 + \cos \theta,
\een
which is the polar equation of a parabola of semi-latus rectum $2$
and origin at its focus.
\end{itemize}

We now turn to the special  case $\gamma=\frac{2}{3}$, that is
$P=-\frac{1}{3} \rho$.
To begin with we assume $k=1$. Now from (\ref{Friedmann})
or  (\ref{Friedmann2})    $\rho_0 \ge 1$,
the pedal equation becomes
\ben
r^2 = p^2 \frac{1}{\sin^2 \alpha},\een
for some constant $\alpha$
which in polar coordinates becomes
\ben
r= a_0\, \e^{\cot \alpha\theta},
\een
where $a_0$ is a constant. This is a logarithmic spiral
unless $\alpha=\frac{\pi}{2}$, in which case it is a circle
of radius $a_0$. Since Friedmann's equation
reads in this case
\ben
(\dot a )^2 = \cot ^2 \alpha,
\een
we see that the scale factor increases or decreases  linearly with time
unless $\alpha = \frac{\pi}{2}$ in which it is constant.

\subsection{A two-component fluid in a closed universe}

For our next example,
consider the locus  of a point at a distance $A$ from the centre of
a circle of radius
$R$. One has
\bea
t&=& R \eta- A \sin \eta,  \label{1.32} \\
a&=& R-A \cos \eta. \label{1.33}
\eea
Note that $\dd t=a \dd \eta$ and thus $\eta$ is conformal time.
Moreover
\ben
\frac{\dd a}{\dd t }= \dot a  = \frac{R \sin \eta}{R- A \cos \eta}.
\een
One thus has
\ben
1+{\dot a }  ^2 = \frac{A^2-R^2 + 2Ra} {a^2} \label{prolate}.
\een
The pedal equation works out to be
\ben
r^2 = A^2 -R^2 + 2Rp,
\een
which may also be verified by a simple geometric argument.
Comparing (\ref{prolate}) with (\ref{Friedmann}) we find that
\ben
\rho =   \frac{2R}{a^3} + \frac{A^2-R^2}{a^4},
\een
which, if $A >R$  is a mixture of dust and radiation with positive
energy density.
The model starts out from a big bang, $a$ increases to a maximum,
and the model ends with a big crunch. This feature may not be so obvious since the time variable transformation (\ref{1.32}) is not invertible when
$A>R$. However, in \cite{CGY}, we have carried out a systematic study which shows that the Friedmann equation (\ref{prolate}) is integrable, in view of
the Chebyshev theorem, to yield a big-bang type periodic solution in cosmic time $t$.

Geometrically we have a {\it prolate
cycloid}.  If $A<R$, we have a {\it curtate cycloid}. The energy density
of the radiation is negative and this gives rise to a bounce and hence
oscillatory behaviour.  This periodic behaviour is clearly seen in (\ref{1.33}) by virtue of the invertible transformation (\ref{1.32}) when $R>A$ but was untouched
in \cite{CGY}.

A more general example is given by the pedal equation of
 an epi- or hypoycloid
\ben
r^2= \frac{p^2}{C} + A^2,
\een
with
\ben
C = \frac{(A+2B)^2}{4B (A+B)}, \label{ratio}
\een
whose parametric equation in
Cartesian coordinates is
\bea
x(\beta)& =& (A+B) \cos \beta -B \cos \left(\frac{A+B }{B} \beta\right),\\
y(\beta) &=& (A+B) \sin  \beta -B \sin \left(\frac{A+B }{B} \beta\right).
\eea

Comparing with (\ref{Friedmann}), we see that
\ben
a^4 \rho +(k-1) a^2
 = \frac{a^2}{C} +A^2,
\een
whence
we have
\ben
\rho = \left( \frac{1}{C} -(1-k) \right) \frac{1} {a^2} + \frac{A^2}{a^4}.  \label{state}
\een
The last term in  (\ref{state}) corresponds to radiation
with a positive pressure,
$\gamma = \frac{4}{3}$. The first term to a fluid with $\gamma=\frac{2}{3}$.
 For particular ratios $A/B$ and hence by (\ref{ratio}) certain values
of $CC$ we have the following special cases

\begin{itemize}

\item $C=\frac{9}{8}$, the Cardoid,\, \qquad $A/B=1$.

\item $C=\frac{4}{3}$, the Nephroid,\, \qquad $A/B=2$.

\item $C=-\frac{1}{8}$,  the Deltoid,\, \qquad $A/B =-3$.  

\item  $C=-\frac{1}{3}$,  the Astroid,\, \qquad $A/B =-4$. 

\end{itemize}

If $k=0$ or $k=1$ , the first two cases correspond
to a fluid with $\gamma =\frac{2}{3}$ with positive
energy density and negative pressure,
the other cases to  negative energy density and positive pressure.

Some other cases are as follows.

\begin{itemize}
\item
\ben
r^2 = \frac{A^2B^2}{p^2} + A^2 - B^2
\een
corresponds to the hyperbola
\ben
\frac{x^2}{A^2} - \frac{y^2}{B^2} =1,
\een
{\it with origin at its  centre}. If $k=1$ this 
has stiff matter  $\gamma =2$ with positive energy density
and radiation whose energy density will be positive
if  $A^2 >B^2$  and negative if   $A^2 <B^2$.
\item
\ben
r^2 = -\frac{A^2B^2}{p^2} + A^2 + B^2
\een
corresponds to the ellipse
\ben
\frac{x^2}{A^2} + \frac{y^2}{B^2} =1,
\een
with origin at its  centre.
If $k=1$ this 
has stiff matter $\gamma =2$  with negative energy density,
in other words, a phantom, and radiation
with positive  energy density.
\item
\ben
r^2 =  \left( \frac{4A^2B^2}{p^2} +4 A^2 \right ) ^{-2}.
\een
This is an ellipse with origin at a focus.
\ben
\frac{x^2}{A^2}  + \frac{y^2}{B^2}=1.\een
If $k=1$,
the energy density is
\ben
\rho = \frac{1}{(2AB)^4} \frac{1}{(1+a^2)^2}. \label{ellipticcatenary}
\een
The curve pursued by the scale factor is called an elliptic catenary.
\item
\ben
r^2 =  \left( \frac{4A^2B^2}{p^2} -4 A^2 \right ) ^{-2}.
\een
This is an hyperbola  {\it but now with origin at a focus}.
\ben
\frac{x^2}{A^2}  - \frac{y^2}{B^2}=1.
\een
If $k=1$, the energy density is
\ben
\rho = \frac{1}{(2AB)^4} \frac{1}{(1-a^2)^2}.
\label{ hyperboliccatenary}
\een
The curve pursued by the scale factor is called a hyperbolic  catenary.

\end{itemize}

\subsection{ $\Lambda{\rm CDM}$ cosmology}

 The presently  favoured, so-called concordance model
 has
$k=0$ and
\ben
 \rho = \frac{1}{a^3} + \frac{\Lambda}{3},
\een
corresponding to dust which is known to be integrable.

The pedal equation is therefore
\bea
r&=& p \sqrt{p^2 \left ( \frac{1}{p^3} + \frac{\Lambda}{3}  + \frac{1}{p ^2 }\right) }\\
&=&  \sqrt{  p  + \frac{\Lambda}{3} p^4   + p ^2 }.
\eea

There is a simple  formula for the scale factor as a function of time
but finding the curve of which it is the roulette  appears to be difficult
since one has to solve a quartic curve for $p$.

 The reader will have noticed that many of the
explicit examples of roulettes giving rise to solutions
of Friedmann's equations descibed above 
have involved ``exotic equations
of state''  with components having  for example negative 
energy densities and/or 
pressures exceeding energy densities in magnitude.
These were  often excluded from   traditional texts on cosmology
on the grounds of lack of physical interest. However 
with the growing  observational support for the presence   of dark energy
the literature has become  a great deal less inhibited 
(see e.g. \cite{Furtherrefs}),
and we have therefore felt justified in including 
as many explicit cases  of the roulette  construction  
as we could find.

\section{Quadratic equation of state}
\setcounter{equation}{0}

It is well known that the flat-universe ($k=0$) Friedmann equation in the linear equation of state case
\be
P=A\rho,
\ee
where $A$ is a constant satisfying the non-phantom condition $A>-1$, may be integrated for arbitrary
values of the cosmological constant $\Lambda$ to allow
big-bang solutions with $a(0)=0$, so that  when $\Lambda=0$ the scale factor $a(t)$ grows following
a power law, when $\Lambda>0$, an exponential law (a dark energy regime), and, when $\Lambda<0$ the
scale factor $a(t)$ oscillates periodically.

In this section, we consider the more general situation \cite{AB1,AB2,SSC} when the equation of state follows the
quadratic law
\be\label{4.2}
P=P_0+A \rho+B \rho^2,
\ee
which may be viewed as a second-order truncation approximation of the general equation of state,
$P=P(\rho)$. We note that cosmologies governed by quadratic and some other more extended forms of equations of state were 
investigated earlier in \cite{NO1,NO2,NO3}. For convenience, we again assume non-phantom condition, $A>-1$, throughout our study.

\subsection{Restrictions to parameters in non-degenerate situations}

Inserting (\ref{4.2}) into the law of energy conservation,
\be\label{4.3}
\dot{\rho}+n(\rho+P)\frac{\dot{a}}a=0,
\ee
we obtain
\be\label{4.4}
J\equiv\int\frac{\dd\rho}{P_0+(1+A)\rho+B\rho^2}=\ln (C a^{-n}),
\ee
where $C>0$ is a constant.

For convenience, use $D$ to denote the discriminant of the quadratic
\be\label{4.5}
Q(\rho)=P_0+(1+A)\rho+B\rho^2.
\ee
That is, $D=(1+A)^2 -4 BP_0$.

For $B\neq0$, there are three subcases to consider.

\begin{enumerate}
\item[(i)] $D=0$. Then
$
J=-\frac1{B(\rho-\rho_0)}$ where $\rho_0=-\frac{1+A}{2B}
$.
Hence, absorbing a possible integrating constant, we have
\be\label{4.6}
-\frac1{B(\rho-\rho_0)}=\ln(Ca^{-n}).
\ee
We are interested in a scale factor $a$ initiating from sufficiently small values and
evolving into arbitrarily large values (`big bang cosmology'). If $B>0$, then $-\rho>0$.
Using this fact and $\rho\geq0$ in (\ref{4.6}), we see that small values of $a$ are not allowed.
If $B<0$, then (\ref{4.6}) assumes the form
\be
\frac1{|B|\rho-\frac12(1+A)}=\ln(Ca^{-n}).
\ee
To allow $a$ to be small, we obtain the restriction $|B|\rho-\frac12(1+A)>0$, which prohibits $a$ to assume large values. In conclusion, the case $D=0$ is ruled out.

\item[(ii)] $D>0$. Then $Q(\rho)$ has two roots:
\be
\rho_{1,2}=\frac{-(1+A)\pm\sqrt{D}}{2B},
\ee
resulting in $J=D^{-\frac12}\ln\left|\frac{\rho-\rho_1}{\rho-\rho_2}\right|$. Hence
\be\label{4.8}
\rho=\frac{\rho_1\pm Ca^{-n\sqrt{D}}\rho_2}{1\pm Ca^{-n\sqrt{D}}},
\ee
where $C>0$ is an updated constant.
Inserting the epoch $a=0$ into (\ref{4.8}), we get
\be\label{4.9}
\rho|_{a=0}=\rho_0=\rho_2=\frac{1+A+\sqrt{D}}{2|B|}>0,\quad B<0.
\ee
In other words we find a sign restriction for $B$. Inserting $a(\infty)=\infty$ into (\ref{4.8}), we get
\be
\rho(\infty)=\rho_1=\frac1{2|B|}([1+A]-\sqrt{[1+A]^2+4|B|P_0})\geq0.
\ee
Therefore we must have
\be
P_0\leq0.
\ee
Using $\rho_1<\rho(t)<\rho_2$ ($\forall t$), we see that (\ref{4.8}) is made precise:
\be\label{4.8p}
\rho=\frac{\rho_1 a^{n\sqrt{D}}+C\rho_2}{a^{n\sqrt{D}}+C}.
\ee
\item[(iii)] $D<0$. Then $Q(\rho)$ has no real roots. We can integrate (\ref{4.4}) to get
\be\label{4.14}
\frac{2B}{\sqrt{-D}|B|}\arctan\frac{|B|}{\sqrt{-D}}\left(2\rho+\frac{1+A}B\right)=\ln(Ca^{-n}).
\ee
Since the right-hand side of (\ref{4.14}) stays bounded, we see
that small values of $a$ are not allowed. Thus the case $D<0$ is also ruled out.

\end{enumerate}

\subsection{Global solutions}

Hence we now aim at integrating the flat-space ($k=0$) Friedmann equation
\be\label{4.10}
\left(\frac{\dot{a}}a\right) ^2=\frac{16\pi G_n}{n(n-1)}\rho+\frac{2\Lambda}{n(n-1)}
\ee
in the case $D>0$ only. For this purpose, we may insert (\ref{4.8p}) into (\ref{4.10}) to get
\be\label{4.12}
\left(\frac{\dot{a}}a\right) ^2=\frac{L_1 a^{n\sqrt{D}}+ L_2}{a^{n\sqrt{D}}+ C},
\ee
where
\be\label{4.17}
L_1=\frac2{n(n-1)}({8\pi G_n}\rho_1+\Lambda),\quad L_2=\frac{2C}{n(n-1)}(8\pi G_n\rho_2+\Lambda).
\ee
Thus, an integration of (\ref{4.12}) gives us
\bea
 t&=&I\equiv\int\sqrt{\frac{a^{n\sqrt{D}}+ C}{L_1 a^{n\sqrt{D}}+ L_2}}\,\frac{\dd a}{a}\nn\\
&=&\frac1{n\sqrt{D}}\int\sqrt{\frac{u+ C}{L_1 u + L_2}}\,\frac{\dd u}u\quad(u=a^{n\sqrt{D}}),
\eea
for which the Chebyshev theorem is not applicable in general.

If $L_1=0$, then $L_2\neq0$ otherwise $\rho_1=\rho_2$ and $D=0$ which is false. Thus
$L_2>L_1=0$. In this case
we can set $u=L_2 v^2-C$ to get
\bea\label{4.18}
t&=&I=\frac2{n\sqrt{D}}\int\frac{L_2 v^2}{L_2 v^2 -C}\,\dd v\quad\quad \nn\\
&=&\frac2{n\sqrt{D}}v+\frac1{n\sqrt{D}}
\sqrt{\frac{C}{L_2}}\ln\left|\frac{v-\sqrt{\frac{C}{L_2}}}{v+\sqrt{\frac{C}{L_2}}}\right|+C_1,
\eea
where $C_1$ is an integration constant and
\be\label{4.19}
v=\sqrt{\frac{a^{n\sqrt{D}}+C}{L_2}}.
\ee
In other words, (\ref{4.18}) and (\ref{4.19}) give us the general solution of the Friedmann equation
when $L_1=0$.

From this solution we see that the initial condition $a=0$, or
\be
v=\sqrt{\frac C{L_2}},
\ee
 cannot be achieved at any finite time but
can only be realized at $t=-\infty$. More precisely, from (\ref{4.18}) and (\ref{4.19}), we may deduce
the property
\be\label{a0}
a(t)=\mbox{O}\left(\e^{\sqrt{\frac{L_2}C}\,t}\right),\quad t\to-\infty.
\ee

Furthermore, from (\ref{4.18}) and (\ref{4.19}), we may also deduce the asymptotic behavior
\be
a(t)=\mbox{O} \left(t^{\frac2{n\sqrt{D}}}\right),\quad t\to\infty.
\ee

Now assume $L_1\neq0$. From (\ref{4.12}) we have $L_1,L_2\geq0$. Hence $L_1>0$. Thus $L_2>L_1>0$.
In this case we can set
\be
v=\sqrt{\frac{u+C}{L_1u+L_2}}\quad\mbox{or}\quad u=-\frac{L_2 v^2-C}{L_1 v^2 -1}=-\frac{\left(\frac{L_2}{L_1}-C\right)}{L_1 v^2-1}-\frac{L_2}{L_1}.
\ee
Consequently, we obtain
\bea\label{4.24}
t&=&I=\frac{2(CL_1-L_2)}{n\sqrt{D}}\int\frac{v^2}{(L_1 v^2-1)(L_2 v^2 -C)}\,\dd v\nn\\
&=&\frac{1}{n\sqrt{D}}\left(\frac1{\sqrt{L_1}}\ln\left|\frac{v+\frac1{\sqrt{L_1}}}{v-\frac1{\sqrt{L_1}}}\right|+
\sqrt{\frac{C}{L_2}}\ln\left|\frac{v-\sqrt{\frac C{L_2}}}{v+\sqrt{\frac C{L_2}}}\right|\right)+C_1,
\eea
where $C_1$ is an integration constant. Since
\be\label{4.25}
v=\sqrt{\frac{a^{n\sqrt{D}}+C}{L_1 a^{n\sqrt{D}}+L_2}},
\ee
we see that $a=0$ corresponds to $v=\sqrt{\frac C{L_2}}$ like before.

From (\ref{4.17}), we have
\be
\frac{L_2}C=\frac2{n(n-1)}(8\pi G_n\rho_2+\Lambda)>\frac2{n(n-1)}(8\pi G_n\rho_1+\Lambda)=L_1,
\ee
which leads to $v>\frac1{\sqrt{L_1}}$. Consequently we see in view of (\ref{4.24}) and (\ref{4.25}) that the scale factor $a$ cannot vanish at any finite time but at $t=-\infty$. More precisely, we have
\be
a(t)=\mbox{O}\left(\e^{\sqrt{\frac{L_2}C}\,t}\right),\quad t\to-\infty,
\ee
which is the same as (\ref{a0}) and may hardly be surprising. Similarly, we have
\be\label{4.29}
a(t)=\mbox{O}\left(\e^{\sqrt{L_1}\, t}\right),\quad t\to\infty.
\ee
This is an important conclusion since we have achieved an exponential asymptotic growth law. It is
worth noting
that the  growth rate $\sqrt{L_1}$ is given by the explicit formula
\bea\label{4.30}
L_1&=&\frac2{n(n-1)}\left(\frac{4\pi G_n}{|B|}\left[(1+A)-\sqrt{(1+A)^2-4|BP_0|}\right]+\Lambda\right),\\
&&\nn\\
&& B<0,\quad P_0\leq0.\nn
\eea

In summary, we can conclude:
\begin{enumerate}
\item[(i)] A nontrivial quadratic equation of state given in (\ref{4.2}) dictates
the necessary and sufficient condition
\bea\label{nsc}
&&P_0\leq0,\quad B<0,\quad (1+A)^2+4|B|P_0>0,\nn\\
&&\\
&& \frac{4\pi G_n}{|B|}\left((1+A)-\sqrt{(1+A)^2-4|BP_0|}\right)+\Lambda\geq0,\nn
\eea
in order to allow the scale factor $a$ to evolve from the epoch $a=0$ to the epoch $a=\infty$.

\item[(ii)] The quadratic equation of state (\ref{4.2}) may never allow a finite initial time at which the
scale factor vanishes.

\item[(iii)] In the critical situation
\be
\frac{4\pi G_n}{|B|}\left((1+A)-\sqrt{(1+A)^2-4|BP_0|}\right)+\Lambda=0,
\ee
the scale factor $a$ grows following a power law. On the other hand, in the non-critical situation
\be
\frac{4\pi G_n}{|B|}\left((1+A)-\sqrt{(1+A)^2-4|BP_0|}\right)+\Lambda>0,
\ee
however, the scale factor $a$ obeys the exponential growth law (\ref{4.29})--(\ref{4.30}). In particular, in this
latter situation, the cosmological constant $\Lambda$ is permitted to assume a suitable negative value when $P_0<0$.
\end{enumerate}

We may compare our results with the $\alpha$-fluid model studied in \cite{Chavanis} for which the
equation of state is
\be\label{Chav1}
P=-(\alpha+1)\rho_{\Lambda}+\alpha\rho-(\alpha+1)\frac{\rho^2}{\rho_P},
\ee
where $\alpha\geq0$ is a parameter, $\rho_P$ is the Planck density, and $\rho_{\Lambda}$ the cosmological density. It is seen that (\ref{Chav1}) is consistent with a
part of the necessary and sufficient condition just derived.  Furthermore, the positivity condition
for the discriminant of (\ref{4.5}), i.e., $D>0$, translates itself into the form
\be\label{Chav2}
\frac{\rho_\Lambda}{\rho_P}<\frac14,
\ee
which does not involve $\alpha$, in particular. Since $\rho_\Lambda$ is of order $10^{-24}$ and
$\rho_P$ of order $10^{99}$ (cf. \cite{Chavanis}), the condition (\ref{Chav2}) is well observed too.

\subsection{Degenerate situation}
In the degenerate situation, $B=0$, (\ref{4.4}) gives us the relation
\be
\left|P_0+(1+A)\rho\right|=C^{1+A} a^{-n(1+A)}.
\ee
Since we are interested in the case when $a$ may assume arbitrarily large values as
$t\to\infty$ and $\rho\geq0$, we again arrive at the necessary condition
\be
P_0\leq0.
\ee
Hence
\be
\rho=\frac1{1+A}\left(C^{1+A} a^{-n(1+A)}+|P_0|\right).
\ee
In other words, the presence of $P_0$ simply and unsurprisingly adds a shift to the cosmological constant.

\subsection{Polytropic equation of state case}

It will also be interesting to extend our study in this section into a slightly more general situation where the equation of state is of a polytropic one given by \cite{Chavanis2,FG}
\be\label{4.38}
P=A\rho+\kappa\rho^{\gamma},\quad A>-1, \quad \gamma\neq1,\quad \gamma\geq0,
\ee
where $\kappa\neq0$ otherwise the equation of state becomes the well-studied linear case.

Inserting (\ref{4.38}) into the equation of energy conservation, we have
\be\label{4.39}
(1+A)\rho^{1-\gamma}+\kappa=C a^{-n(1+A)(1-\gamma)},
\ee
where $C>0$ is an integration constant.

\begin{enumerate}
\item[(i)] $\gamma<1$. Then (\ref{4.39}) indicates that the epoch $a=0$ gives rise to 
the initial value $\rho_0=\infty$
and the epoch $a=\infty$ gives rise to the condition $\kappa<0$ and the limiting value
\be
\rho_\infty=\left(\frac{|\kappa|}{1+A}\right)^{\frac1{1-\gamma}}.
\ee
\item[(ii)] $\gamma>1$. Then (\ref{4.39}) implies that the epoch $a=0$ gives rise to 
the condition $\kappa<0$ again and
the initial value 
\be
\rho_0=\left(\frac{|\kappa|}{1+A}\right)^{\frac1{1-\gamma}},
\ee
and the epoch $a=\infty$ gives rise to the limiting value $\rho_\infty=0$.
\end{enumerate}

Note that the necessary condition $\kappa<0$ is analogous to the necessary condition $B<0$ stated in
(\ref{nsc}) for the quadratic equation of state problem.

Applying (\ref{4.39}) in the flat-space Friedmann equation with $\Lambda=0$ for simplicity, we arrive
at the integral
\be
4\left(\frac{\pi G_n}{n(n-1)}\right)^{\frac12}(1+A)^{-\frac1{2(1-\gamma)}}\,t=I=\int a^{-1}\left(Ca^{-n(1+A)(1-\gamma)}+|\kappa|\right)^{-\frac1{2(1-\gamma)}}\,\dd a.
\ee
In view of the Chebyshev theorem, we know that the integral $I$ is elementary when $A$ and $\gamma$
are arbitrary rational numbers. Here we omit presenting the solutions.

\section{The Randall--Sundrum II cosmology}
\setcounter{equation}{0}

In this section we consider the braneworld cosmological Friedmann equation \cite{Ge}
\be\label{5.1}
H^2+\frac k{a^2}=\alpha_1\rho +\alpha_2\rho^2+\lm
\ee
in the Randall--Sundrum II universe \cite{RS,SWC,CC} where
\be
\alpha_1=\frac{16\pi G_n}{n(n-1)},\quad \alpha_2=\left(\frac{16\pi G_{n+1}}n\right)^2,\quad\lm=\frac{2\Lambda}{n(n-1)}
\ee
are specific physical parameters.

\subsection{Linear equation of state case}
In view of the linear equation of state
\be\label{5.3}
P=A\rho
\ee
and the law of energy conservation (\ref{4.3}), we have
\be\label{5.4}
\rho=Ca^{-n(1+A)},\quad 1+A>0,\quad C>0.
\ee

Inserting (\ref{5.4}) into (\ref{5.1}) and restricting to the flat-space case $k=0$, we easily obtain the integration
\bea\label{5.5}
t+C_1&=&I=\int\frac{\dd a}{a\sqrt{\beta_1 a^{-n(1+A)}+\beta_2 a^{-2n(1+A)}+\lm}}\nn\\
&=&\int\frac{a^{n(1+A)-1}\dd a}{\sqrt{\lm a^{2n(1+A)}+\beta_1 a^{n(1+A)}+\beta_2}}\nn\\
&=&\frac1{n(1+A)}\int\frac{\dd u}{\sqrt{\lm u^2+\beta_1 u+\beta_2}}\quad \quad(u=a^{n(1+A)})\nn\\
&=&\left\{\begin{array}{lll}\frac2{n(1+A)\beta_1}\sqrt{\beta_1 a^{n(1+A)}+\beta_2}, &\lm=0,\\
\frac1{n(1+A)\sqrt{\lm}}\ln\left(\frac{\beta_1}{2\lm}+a^{n(1+A)}+\sqrt{a^{2n(1+A)}+\frac{\beta_1}\lm a^{n(1+A)}+\frac{\beta_2}\lm}\right), &\lm>0,\\
\frac1{n(1+A)\sqrt{|\lm|}}\arctan\left(\frac{a^{n(1+A)}-\frac{\beta_1}{2|\lm|}}{\sqrt{-a^{2n(1+A)}+\frac1{|\lm|}(\beta_1 a^{n(1+A)}
+\beta_2)}}\right), & \lm<0,\end{array}\right.\nn\\
\eea
since it is of the Chebyshev form, where $\beta_1=C\alpha_1$ and $\beta_2=C^2\alpha_2$ and $C_1$ is
an integration constant which may always be chosen to allow the big-bang initial condition $a(0)=0$.
Indeed, from (\ref{5.5}), we have
\be
C_1=\left\{\begin{array}{lll}\frac{2\sqrt{\beta_2}}{n(1+A)\beta_2},&\lm=0,\\
\frac1{n(1+A)\sqrt{\lm}}\ln\left(\frac{\beta_1}{2\lm}+\sqrt{\frac{\beta_2}\lm}\right), &\lm>0,\\
-\frac1{n(1+A)\sqrt{|\lm|}}\arctan\left(\frac{\beta_1}{2\sqrt{|\lm|\beta_2}}\right),&\lm<0.\end{array}
\right.
\ee

From the above results we deduce, as $t\to\infty$, the growth laws
\be
a(t)=\mbox{O}\left(t^{\frac2{n(1+A)}}\right),\quad \lm=0;\quad a(t)=\mbox{O}\left(\e^{\sqrt{\lm}t}\right),\quad\lm>0.
\ee
These two situations are analogous to those in the classical situation \cite{CGLY,SK}.

However, it may be a surprise to see, when $\lm<0$, the solution will be of a finite lifespan or periodic, instead, depending on how far the parameter $A$ is away from
the phantom divide line \cite{xV,xMc,xC-L,xN-P,xG-C}, $A=-1$. To see this fact more transparently, we
rewrite the Friedmann equation as
\be
\left(\frac{\dd u}{\dd t}\right)^2=n^2(1+A)^2{|\lm|}(u_1-u)(u+u_2),
\ee
where
\be
u_1=\frac1{2|\lm|}\left(\beta_1+\sqrt{\beta_1^2+4\beta_2|\lm|}\right),\quad
u_2=\frac1{2|\lm|}\left(\sqrt{\beta_1^2+4\beta_2|\lm|}-\beta_1\right),
\ee
and $u=a^{n(1+A)}$. Using $u(0)=0$, we solve
\be\label{5.10}
\frac{\dd u}{\dd t}=\alpha\sqrt{(u_1-u)(u+u_2)}, \quad \alpha={n(1+A)\sqrt{|\lm|}},\quad t>0,
\ee
to get
\be
\alpha t=\arctan\left(\frac{u-\frac12(u_1-u_2)}{\sqrt{(u_1-u)(u+u_2)}}\right)+\arctan\left(\frac12\frac{(u_1-u_2)}{\sqrt{u_1 u_2}}\right),\quad t>0.
\ee
The above relation remains valid until $t=t_0>0$ (say) when $u=u_1$. Thus
\be
t_0=\frac1\alpha\left(\frac\pi2+\arctan\left[\frac12\frac{(u_1-u_2)}{\sqrt{u_1 u_2}}\right]\right).
\ee
Beyond $t_0$ the equation (\ref{5.10}) is invalid. Instead, $u$ reaches a `stagnation' point, where
$\dot{u}(t_0)=0$ and $u(t_0)=u_1$, and starts to descend. Hence it must evolve through another branch of the Friedmann equation,
\be\label{5.13}
\frac{\dd u}{\dd t}=-\alpha\sqrt{(u_1-u)(u+u_2)}, \quad \alpha={n(1+A)\sqrt{|\lm|}},\quad t>t_0,
\ee
which may be integrated similarly. Actually, since both (\ref{5.10}) and (\ref{5.13}) are autonomous,
the solution of (\ref{5.13}) satisfying $u(t_0)=u_1$ may simply be obtained from that of (\ref{5.10}) through replacing
the variable $t$ with $t\in [0,t_0]$ for the solution of (\ref{5.10})
by $2t_0-t$ with $t\in [t_0,2t_0]$. In particular, $u(2t_0)=0$.

\subsection{Solutions of finite lifespans}
We now show that the solution  ceases to exist beyond $2t_0$ when
\be\label{xA}
A\geq -1+\frac1n.
\ee
In fact, from (\ref{5.13}), we have
\be
\lim_{t\to (2t_0)^-}\dot{u}(t)=-\alpha\sqrt{u_1 u_2}<0.
\ee
From the relation between the scale factor $a$ and the solution $u$ we find
\be\label{5.16}
\dot{a}(t)=\frac1{n(1+A)} (u(t))^{-\left(1-\frac1{n(1+A)}\right)}\dot{u}(t),
\ee
so that
\be\label{5.17}
\lim_{t\to(2t_0)^-} \dot{a}(t)=-\infty,\quad A>-1+\frac1n;\quad \lim_{t\to(2t_0)^-}\dot{a}(t)=-\alpha\sqrt{u_1u_2}<0,\quad A=-1+\frac1n.
\ee
On the other hand, if the solution would exist beyond $2t_0$, we should have $\dot{a}(t)\geq0$ when
$t$ passes $2t_0$ where $a=0$. Thus (\ref{5.17}) indicates that $\dot{a}(t)$ would suffer from a discontinuity at $t=2t_0$. In other words, the solution to the Friedmann equation cannot exist
beyond $2t_0$ which may well be called the lifespan of the solution.

If the coupling constant $A$ satisfies
\be\label{5.18}
A<-1+\frac1n,
\ee
then (\ref{5.16}) implies that $\dot{a}(t)\to0$ as $t\to (2t_0)^-$. Thus we may obtain
the solution of the Friedmann equation over $[2t_0,4t_0]$ by shifting the
solution over $[0,2t_0]$ obtained above and maintain the continuity of $\dot{a}(t)$ at $t=2t_0$. Such
a procedure can be repeated so that a periodic solution of period $2t_0$ is constructed.

Thus, we conclude that, when the cosmological constant is negative, the solution $a$ of the Friedmann equation evolving from the
initial condition $a(0)=0$ has either a finite lifespan $T=2t_0$ or is of period $T=2t_0$, according to whether the coupling constant $A$
fulfills the condition (\ref{xA}) or (\ref{5.18}).

For later convenience, we document the lifespan
or period $T$ here explicitly in terms of the original physical parameters:
\be
T=\frac1{1+A}\sqrt{\frac{n-1}{2n|\Lambda|}}\left(\pi+2\arctan\left[\frac{G_n}{2G_{n+1}}\sqrt{\frac n{2(n-1)|\Lambda|}}\right]\right),\quad\Lambda<0.
\ee

Thus we have unveiled a peculiar situation, $A\geq -1+\frac1n$, when the solution has only a finite
lifespan. This situation is in sharp contrast with the classical case with $\Lambda<0$ that the
big-bang solution is periodic for any $A>-1$.

To end this subsection, it will be instructive to check that, subject to the equation of state
(\ref{5.3}), the phantom world is still spelled out by the
condition $1+A<0$ as in the classical situation. In fact, for $1+A\neq0$, the
energy conservation law gives us the relation $\rho=Ca^{-n(1+A)}$ as before where $C>0$ is a constant. Inserting this into the Friedmann equation (\ref{5.1}) with $k=0,\lambda=0$, we have the integration
\be
t=I=\int\frac{\dd a}{a\sqrt{C\alpha_1 a^{-n(1+A)}+C^2\alpha_2 a^{-2n(1+A)}}}.
\ee
Hence the scale factor $a=a(t)$ is implicitly given by
\be
\frac{\alpha_1}C a^{n(1+A)}+\alpha_2=\left(\sqrt{\left(\frac{\alpha_1}C\right)a^{n(1+A)}(0)+\alpha_2}+\frac12n(1+A)\alpha_1 t\right)^2,\quad t>0.
\ee
Consequently we see that the initial condition $a(0)=0$ is permitted when $1+A>0$ so that $a=a(t)$ 
grows following the law
\be
a(t)=\mbox{O}\left(t^{\frac2{n(1+A)}}\right),\quad t\to\infty.
\ee
If $1+A<0$, however, the initial condition $a(0)=0$ is not permitted 
and the Big Rip \cite{EMM,CKW} happens at 
\be
t_0=\frac{2}{n|1+A|\alpha_1}\sqrt{\left(\frac{\alpha_1}C\right)a^{n(1+A)}(0)+\alpha_2}.
\ee
In particular, $A=-1$ still corresponds to the phantom divide line \cite{xV,xMc,xC-L,xN-P,xG-C} indeed as in the classical situation, as anticipated.

In \cite{HM-R}, the following slightly more general Friedmann equation 
\be
H^2=\alpha\rho+\beta\rho^2+\frac\gamma{a^4}
\ee
is considered ($k=0,\Lambda=0$), where $\alpha,\beta,\gamma>0$ are constants. Although this equation
cannot be integrated, some asymptotic analysis establishes that any solution with $a(\infty)=\infty$
grows according to the law $a(t)=\mbox{O}\left(t^{\frac2{n(1+A)}}\right)$ as $t\to\infty$, as before, subject to the
equation of state $P=A\rho$ ($A>-1$).

\subsection{Chaplygin fluid case}

Consider the equation of state
\be\label{5.20}
P=A\rho-\frac{B}\rho,\quad A>-1,\quad B>0,
\ee
of a generalized Chaplyin fluid model \cite{ST,CP,ABL}. Inserting (\ref{5.20}) into the law of energy conservation, (\ref{4.3}), we obtain
\be\label{5.21}
\rho=\left(Ca^{-2n(1+A)}+\frac B{1+A}\right)^{\frac12},
\ee
where $C>0$ is an integration constant. Substituting (\ref{5.21}) into the braneworld Friedmann equation (\ref{5.1}) with $k=0$ and $\lm=0$ and integrating, we arrive at
\be\label{5.22}
t=I\equiv\int\left(\alpha_1\left[Ca^{-2n(1+A)}+b\right]^{\frac12}+\alpha_2\left[Ca^{-2n(1+A)}+b\right]\right)^{-\frac12}\frac{\dd a}a,\quad b=\frac B{1+A}.
\ee

To proceed, we note that (\ref{5.21}) gives us
\be\label{5.23}
a=C^{\frac1{2n(1+A)}}(\rho^2-b)^{-\frac1{2n(1+A)}},\quad \rho^2> b.
\ee
Combining (\ref{5.22}) and (\ref{5.23}), we have
\be\label{5.24}
I=-\frac1{n(1+A)}\int\frac1{(\rho^2-b)}\sqrt{\frac{\rho}{\alpha_1+\alpha_2\rho}}\,\dd\rho,
\ee
which is not of the Chebyshev type. Nevertheless, set
\be
u=\sqrt{\frac{\rho}{\alpha_1+\alpha_2\rho}}.
\ee
Then $0\leq u<\frac1{\sqrt{\alpha_2}}$ and
\be
\rho=\frac{\alpha_1 u^2}{1-\alpha_2 u^2},
\ee
which renders the integral (\ref{5.24}) into a rational one,
\be
I=-\frac{2}{\alpha_1 n(1+A)}\int\frac{u^2}{u^4- b{\alpha_1^{-2}}(1-\alpha_2 u^2)^2}\,\dd u,
\ee
which can be integrated separately in the parameter regimes
\be
\alpha_2>\frac{\alpha_1}{\sqrt{b}},\quad \alpha_2=\frac{\alpha_1}{\sqrt{b}},\quad
\alpha_2<\frac{\alpha_1}{\sqrt{b}},
\ee
respectively.

We begin by displaying the result in the first regime as follows:
\bea\label{5.29}
I&=&\frac1{n(1+A)\sqrt{b}}\left\{\left(\alpha_2+\frac{\alpha_1}{\sqrt{b}}\right)^{-\frac12}\arctanh\left(\left(\alpha_2+\frac{\alpha_1}{\sqrt{b}} \right)^{\frac12} u\right)\right.\nn\\
&&\quad\quad -
\left.\left(\alpha_2-\frac{\alpha_1}{\sqrt{b}}\right)^{-\frac12}\arctanh\left(\left(\alpha_2-\frac{\alpha_1}{\sqrt{b}}\right)^{\frac12} u\right)\right\}+C_1,\quad \alpha_2>\frac{\alpha_1}{\sqrt{b}},
\eea
where $C_1$ is an integration constant. It is clear that we may choose $C_1$ suitably to allow
the big-bang initial condition $a(0)=0$ or $u(0)=\frac1{\sqrt{\alpha_2}}$. On the other hand,
the growth condition $a(\infty)=\infty$ gives rise to $\rho(\infty)=\sqrt{b}$. Consequently,
we arrive at
\be\label{5.30}
u(\infty)=\left(\alpha_2+\frac{\alpha_1}{\sqrt{b}}\right)^{-\frac12},
\ee
as anticipated. In other words, the first term on the right-hand side of (\ref{5.29}) dominates
as $t\to\infty$, which leads to the following explicit asymptotic estimate:
\be\label{5.31}
a(t)=\mbox{O}\left(\e^{\sigma t}\right),\quad t\to\infty,
\ee
where
\be\label{5.32}
\sigma={\sqrt{b}}\left(\alpha_2+\frac{\alpha_1}{\sqrt{b}}\right)^{\frac12}
=\sqrt{\frac B{1+A}}\left(\left[\frac{16\pi G_{n+1}}n\right]^2+\frac{16\pi G_n}{n(n-1)}\sqrt{\frac{1+A}B}\right)^{\frac12},
\ee
under the condition
\be
\alpha_1<\alpha_2\sqrt{b}\quad\mbox{or}\quad G_n<16\pi\left(1-\frac1n\right)G^2_{n+1}\sqrt{\frac B{1+A}}.
\ee

Likewise, in the third regime, we have
\bea\label{5.34}
\alpha_1 n(1+A)I
&=&\frac{\alpha_1}{\sqrt{b}}\left(\frac{\alpha_1}{\sqrt{b}}+\alpha_2\right)^{-\frac12}\arctanh\left(\left(\frac{\alpha_1}{\sqrt{b}}+\alpha_2\right)^{\frac12}u\right)\nn\\
&&-\frac{\alpha_1}{\sqrt{b}}\left(\frac{\alpha_1}{\sqrt{b}}-\alpha_2\right)^{-\frac12}\arctan\left(\left(\frac{\alpha_1}{\sqrt{b}}-\alpha_2\right)^{\frac12}u\right)+C_1,\nn\\
&& \mbox{where }\alpha_2<\frac{\alpha_1}{\sqrt{b}}\quad \mbox{or }16\pi\left(1-\frac1n\right)G^2_{n+1}\sqrt{\frac B{1+A}}<G_n.
\eea

With (\ref{5.34}), we have the same asymptotic state (\ref{5.30}) and we see that (\ref{5.31}) and
(\ref{5.32}) still hold.

We omit here the critical case $\sqrt{b}\alpha_2=\alpha_1$.

It is interesting to note the different ways the Newton constants $G_n$ and $G_{n+1}$ and the constants $A$
and $B$ make their contributions to the dark energy, $\sigma$, as expressed in (\ref{5.32}).

It will also be enlightening to compare (\ref{5.31})--(\ref{5.32}) with the formulas \cite{CGY}
\be
\quad \sigma=4\left(\frac{\pi G_n}{n(n-1)}\right)^{\frac12}\left(\frac B{1+A}\right)^{\frac14},
\ee
obtained for the big-bang solution of the classical Friedmann equation
\be
H^2=\frac{16\pi G_n}{n(n-1)}\rho,
\ee
which may simply be obtained from (\ref{5.32}) by setting $G_{n+1}=0$.

\section{Universal growth law for the extended Chaplygin universe}
\setcounter{equation}{0}

Recall that the generalized Chaplygin fluid is governed by the equation of state \cite{ST,CP,ABL}
\be\label{6.1}
P=A\rho-\frac B{\rho^\alpha},\quad A>-1,\quad B>0,\quad 0\leq\alpha\leq 1,
\ee
whose cosmological implication, especially its relevance to dark energy problem, has been studied
extensively. It is direct to see from (\ref{6.1}) that there holds the scale factor and energy density relation
\be\label{6.2}
\rho^{\alpha+1}=Ca^{-n(1+A)(\alpha+1)}+\frac B{1+A}.
\ee

In view of the Chebyshev theorem, we know that the Friedmann equation with $k=0$ and
$\Lambda=0$ is integrable
\cite{CGY} when $\alpha$ is rational. Nevertheless, for any real $\alpha$, we derived in \cite{CGY}
the following explicit universal growth law:
\be\label{6.3}
a(t)\sim C_0\e^{\sigma t},\quad t\to\infty;\quad \sigma=4\left(\frac{\pi G_n}{n(n-1)}\right)^{\frac12}
\left(\frac B{1+A}\right)^{\frac1{2(\alpha+1)}}.
\ee

Inserting (\ref{6.3}) into (\ref{6.2}), we have
\be\label{6.4}
\rho(t)\to\rho_\infty=\left(\frac B{1+A}\right)^{\frac1{\alpha+1}},\quad t\to\infty.
\ee

Unsuprisingly, $\rho_\infty$ is the only positive root of the function
\be
h(\rho)=P+\rho=\rho+A\rho-\frac B{\rho^\alpha}.
\ee

Suggested by the above results, we establish in this section a universal growth law in much extended situations, without resorting to an integration of the Friedmann equation.

\subsection{Extended Chaplygin fluids}

In \cite{PK1,PK2} and \cite{KKPMP}, the extended Chaplygin fluid model given by the equation of state
\be
P=\sum_{l=1}^M A_l\rho^l -\frac B{\rho^\alpha}
\ee
is studied in the specific situations when $A_l=A>0$ ($l=1,\dots,M$) and $A_l=\frac1l$ ($l=1,\dots,M$), respectively, by numerical means. Motivated by their work, here we consider the extended
Chaplygin model governed by the following general equation of state:
\be
P=f(\rho)-\sum_{m=1}^N \frac{B_m}{\rho^{\alpha_m}},\quad \rho>0,
\ee
where $B_1,\dots,B_N>0$ and $\alpha_1,\dots,\alpha_N\geq0$ are constants
 and $f$ is an analytic function satisfying
\be
f(0)=0;\quad f'(\rho)>-1,\quad\rho>0;\quad  \rho+f(\rho)\to\infty\quad\mbox{as }\rho\to\infty.
\ee

A direct consequence of the above assumption on $f$ is that
 the function
\be\label{6.8}
h(\rho)\equiv \rho+f(\rho)-\sum_{m=1}^N\frac {B_m}{\rho^{\alpha_m}}
\ee
has exactly one positive root, say $\rho_\infty>0$, since $h(\rho)$ strictly increases in $\rho>0$.

It is clear that our assumption on $f$, motivated from the
results in \cite{CGY} and discussed above, covers all the examples considered in \cite{PK1,PK2,KKPMP}.

\subsection{Universal growth law}

Rewrite the function $h$ in (\ref{6.8}) in the form
\be
h(\rho)=(\rho-\rho_\infty) h_1(\rho).
\ee
Since $h(\rho)<0$ for $\rho>0$ small, we see that $h_1(\rho)>0$ for all $\rho>0$. Inserting this into
the law of energy conservation (\ref{4.3}), we have
\be\label{6.11}
\int_{\rho_0}^{\rho}\frac{\dd r}{h_1(r)(r-\rho_\infty)}=\ln \left(\frac{a(t_0)}a\right)^n,
\ee
where $\rho_0=\rho(t_0)>0$ and $t_0$ is an initial time when $a(t_0)>0$
and  $\rho=\rho(t),a=a(t)$ with $t\geq t_0$. Note that we are dealing with an expanding universe such that $\rho_0\geq\rho>\rho_\infty$.

In order to extract information from (\ref{6.11}), we rewrite the integrand of it as
\bea\label{6.12}
\frac1{h_1(r)(r-\rho_\infty)}&=&\frac1{h_1(\rho_\infty)(r-\rho_\infty)}+\frac1{r-\rho_\infty}\left(\frac1{h_1(r)}
-\frac1{h_1(\rho_\infty)}\right)\nn\\
&\equiv&\frac1{h_1(\rho_\infty)(r-\rho_\infty)}+q(r),\quad r>0,
\eea
where $q(r)$ is a smooth function in $r>0$. Inserting (\ref{6.12}) into (\ref{6.11}), we arrive at
\be\label{6.13}
\frac1{h_1(\rho_\infty)}\ln\left|{\rho-\rho_\infty}\right|
=Q(\rho)+\ln \left(a^{-n}\right),
\ee
where
\be\label{6.14}
Q(\rho)=\frac1{h_1(\rho_\infty)}\ln|\rho_0-\rho_\infty|+n\ln \left(a(t_0)\right)-\int_{\rho_0}^\rho q(r)\,\dd r,\quad \rho_\infty<\rho\leq\rho_0,
\ee
which is a bounded function. Combining (\ref{6.13}) and (\ref{6.14}), we have
\be\label{6.15}
\rho=\rho_\infty+\e^{h_1(\rho_\infty)Q(\rho)} a^{-nh_1(\rho_\infty)}.
\ee
Inserting (\ref{6.15}) into the Friedmann equation (\ref{4.10}), we get
\be\label{6.16}
\left(\frac{\dot{a}}{a}\right)^2=\lm_0+\alpha\,\e^{h_1(\rho_\infty)Q(\rho)} a^{-nh_1(\rho_\infty)},\quad \alpha=\frac{16\pi G_n}{n(n-1)},
\ee
where we set
\be\label{6.17}
\lm_0=\frac{16\pi G_n}{n(n-1)}\rho_\infty+\frac{2\Lambda}{n(n-1)}>0.
\ee
In view of (\ref{6.16}) and (\ref{6.17}), we obtain the integration
\bea\label{6.18}
t&=&\int\frac{\dd a}{a\sqrt{\lm_0+\alpha\,\e^{h_1(\rho_\infty)Q(\rho)} a^{-nh_1(\rho_\infty)}}}\nn\\
&=&\int\frac{\dd a}{\sqrt{\lm_0}\,a}+\int\left(\frac{1}{a\sqrt{\lm_0+\alpha\,\e^{h_1(\rho_\infty)Q(\rho)} a^{-nh_1(\rho_\infty)}}}-\frac{1}{\sqrt{\lm_0}\,a}\right)\,\dd a\nn\\
&=&\frac1{\sqrt{\lm_0}}\ln a+R(a),
\eea
where
\bea\label{6.19}
R(a)&=&\int\left(\frac{1}{a\sqrt{\lm_0+\alpha\,\e^{h_1(\rho_\infty)Q(\rho)} a^{-nh_1(\rho_\infty)}}}-\frac{1}{\sqrt{\lm_0}\,a}\right)\,\dd a\nn\\
&=&
-\frac12 \int\frac{\alpha\e^{h_1(\rho_\infty)Q(\rho)} a^{-1-nh_1(\rho_\infty)}}{\left(\sqrt{\lm_0+\alpha\xi(a)\e^{h_1(\rho_\infty)Q(\rho)} a^{-nh_1(\rho_\infty)}}\right)^3}\,\dd a,\quad
\xi(a)\in [0,1],
\eea
which is bounded for $a$ near $\infty$ since $h_1(\rho_\infty)>0$. Consequently, in view of (\ref{6.18}) and (\ref{6.19}), we deduce the universal growth law for the scale factor $a$ as follows:
\be\label{6.20}
a(t)=\mbox{O}\left( \e^{\sqrt{\lm_0}t}\right),\quad t\to\infty,
\ee
where $\lm_0$ is given in (\ref{6.17}).

\subsection{Examples}

It will be enlightening to display a few special examples which allow explicit calculations of $\rho_\infty$,  and hence, $\lm_0$.

\begin{enumerate}
\item[(i)] For the classical generalized Chaplygin model (\ref{6.1}), we see that $\rho_\infty$ is given by (\ref{6.4}).
Hence
\be\label{6.21}
\lm_0=\frac{16\pi G_n}{n(n-1)}\left(\frac B{1+A}\right)^{\frac1{\alpha+1}}+\frac{2\Lambda}{n(n-1)},
\ee
which generalizes the corresponding formula obtained in \cite{CGY}.

\item[(ii)] Consider the extended Chaplygin fluid governed by the equation of state
\be
P=A_1\rho +A_3\rho^3-\frac B\rho.
\ee
It is direct to see that
\be
\rho_\infty=\left(\frac1{2A_3}\left[\sqrt{(1+A_1)^2+4A_3 B}-(1+A_1)\right]\right)^{\frac12}.
\ee
Consequently we have
\be\label{6.24}
\lm_0=\frac{16\pi G_n}{n(n-1)}\left(\frac{2B}{(1+A_1)+\sqrt{(1+A_1)^2+4A_3 B}}\right)^{\frac12}+
\frac{2\Lambda}{n(n-1)}.
\ee
In the limit $A_3\to 0$, (\ref{6.24}) recovers (\ref{6.21}) for $\alpha=1$.

\item[(iii)] As another simple example, let the equation of state be
\be
P=A\rho-B_0-\frac{B_1}\rho.
\ee
Then we have
\be
\rho_\infty=\frac{B_0+\sqrt{B_0^2+4B_1(1+A)}}{2(1+A)},
\ee
which allows us to return to (\ref{6.21}) with $\alpha=1$ again in the limit $B_0\to0$.
\end{enumerate}

There are other explicitly solvable cases of interest as well. For example, we consider the equation
of state
\be
P=A\rho-\frac{B_1}{\rho^{\alpha_1}}-\frac{B_2}{\rho^{\alpha_2}},\quad \alpha_2>\alpha_1\geq0.
\ee
To solve the associated equation
\be
(1+A)\rho-\frac{B_1}{\rho^{\alpha_1}}-\frac{B_2}{\rho^{\alpha_2}}=0,
\ee
we require $1+\alpha_2=2(\alpha_2-\alpha_1)$ or $\alpha_2=1+2\alpha_1$. Hence $\alpha_1\geq0$ is
allowed to be arbitrary, which contains the case (iii) above as a special example where $\alpha_1=0$.
In other words, we find an explicit solvable general situation with the equation of state
\be\label{6.29}
P=A\rho-\frac{B_1}{\rho^\alpha}-\frac{B_2}{\rho^{1+2\alpha}},\quad\alpha\geq0.
\ee
With (\ref{6.29}), we obtain $\rho_\infty$ (hence $\lm_0$) explicitly:
\be
\rho_\infty=\left(\frac{B_1+\sqrt{B_1^2+4(1+A)B_2}}{2(1+A)}\right)^{\frac1{1+\alpha}}.
\ee
 
Similarly, with
\be
P=A_1\rho+A_2\rho^{2+\alpha}-\frac B{\rho^\alpha},\quad\alpha\geq0,
\ee
we have
\be
\rho_\infty=\left(\frac{\sqrt{(1+A_1)^2+4A_2 B}-(1+A_1)}{2A_2}\right)^{\frac1{1+\alpha}},
\ee
which covers (ii) with $\alpha=1$.

\section{Conclusions}

Our results in this work are summarized as follows.

\begin{enumerate}
\item[(i)] {\em Friedmann equation and roulettes.}
We have shown that  every solution of Friedmann equation
admits a representation as a roulette, that is the  locus of a point in the
plane of a curve which rolls  without  slipping on a straight line.
A method is given   for  finding  the curve.
A well-known  example is the scale factor
of  closed universe containing pressure free matter.
The scale factor is a  cycloid, i.e. the
locus of a point on the circumference of a  circle, which rolls without
slipping on a straight line.
Adding  radiation, one finds that curve is still a circle
but  the point moves
outside its  circumference.
Many other explicit examples are given, some involving
exotic equations of state which have been of current
interest to cosmologists  in connection with dark energy and the
pre-inflationary universe.

The Friedmann equation, or its equivalent,
also arises in other contexts in   physics not
directly connected with cosmology such as central orbit
problems and geometric optics. Examples
of the use of roulettes  are in this wider setting are
given.

\item[(ii)] {\em Quadratic equation of state.}  
Although the flat-space  Friedmann equation is not in a binomial form so that the Chebyshev theorem is
applicable,  a necessary and sufficient condition is unveiled for the
coefficients of the equation of state to allow the scale factor $a$ to evolve from zero to infinity.
A distinguishing characteristic is that the presence of the quadratic term in the equation of state
prohibits $a$ to vanish at a finite initial time. In other words, the epoch $a=0$ may only happen
at $t=-\infty$. In the general non-critical situations, $a$ grows exponentially even when the cosmological
constant $\Lambda$ assumes a negative value above a specific critical level.

\item[(iii)] {\em Randall--Sundrum II universe.}
In the Randall--Sundrum II universe, there are two interesting flat-space cases: When the equation of state is linear, it is shown that
the scale factor $a$ grows following a power law or an exponential law according to whether the cosmological
constant $\Lambda$ is zero or positive, as in the classical situation, but when $\Lambda$
is negative, however, a new phenomenon occurs so that in an explicitly given regime away from the phantom divide line, the big-bang solution satisfying $a(0)=0$ has only a finite lifespan, which never happens in
the classical situation, although near the phantom divide line the solution is periodic, as
in the classical situation; when the equation of state is
that of a Chaplygin fluid, we explicitly derive the  big-bang solution and describe its exponential
growth pattern in terms of various coupling parameters, when $\Lambda=0$. We see as in the classical
Chaplygin fluid model case that a small amount of exotic nonlinear matter would lead to a large amount
of presence of dark energy near the phantom divide line, as realized by the exponential growth
rate formula.

\item[(iv)] {\em Universal growth formula.} For the extended Chaplygin fluid universe for which 
the equation of state is decomposed into the sum of a positive analytic function
and a negative inverse power function, an integration is impossible in general.
To tackle the problem, an analytic method is introduce which enables us to extract
 all the key parameters that give rise to a universal formula for the exponential growth rate
of the scale factor. Such a formula covers all known formulas derived in concrete situations.

\end{enumerate}

\medskip

The research of Chen was supported in part by Henan Basic Science and Frontier
Technology Program Funds under Grant No. 142300410110. Yang was partially supported
by National Natural Science Foundation of China under Grant No. 11471100.

\small{

\section{Appendix:  Pedal equations and their inversion}
\setcounter{equation}{0}

If  a curve in the plane  is given in polar coordinates $(r,\phi)$ by
\ben
u= \frac{1}{r}= u(\phi),
\een
and if $p$ is  the perpendicular  distance from the origin to the
tangent at the point $(r,\phi)$,
one has
\ben
\frac{1}{p^2}= \left(\frac{\dd u}{\dd\phi}\right )^2   + u^2.
\een
Conversely, if we are given the pedal equation of a curve in the form
\ben
\frac{1}{p^2}= P(u),
\een
the problem of finding the polar equation of the
curve amounts to solving the differential equation
\ben
\frac{1}{p^2}= \left (\frac{\dd u}{\dd\phi}\right )^2   + u^2 = P(u). \label{8pedal}
\een

Equation (\ref{8pedal}) arises in many other contexts. Here are two examples.
\begin{itemize}
\item If a particle moves in a spherically symmetric potential $V(\frac1r)$
per unit mass and   has conserved energy per unit mass
$\cE$ and conserved angular momentum per unit mass $h$ then its orbit
satisfies (\ref{8pedal})   with
\ben
P(u) = \frac{2}{h^2} \left ( \cE - V\left(\frac{1}{u}\right) \right).
\een
\item The scale factor $a(t)$ of Friedmann--Lemaitre universe with energy density
$\rho(a)$ (including any contribution due to a cosmological term)  satisfies
\ben
\frac{\dot a ^2 }{a^2} + \frac{1}{a^2}  =
4 \pi G  \rho(a) + \frac{(1-k)}{ a^2},
\een
which may be brought to the form  (\ref{8pedal}) with $u=\frac{1}{a}$,
$\phi= \dd \eta = a \dd t$,
and
\ben
P(u) =  4 \pi G a^2 \rho(a) + (1-k) a^2.
\een
\end{itemize}

\subsection{Affine and rescaling of angle}

Suppose $u_1=f(\phi) $ satisfies
\ben
\left(\frac{\dd u_1}{\dd\phi}\right )^2 + u_1^2 = P_1(u_1).
\een

Let $u_2= A f(\lambda \phi) + B $. Then
\ben
\left(\frac{\dd u_2}{\dd\phi}\right )^2 + u_2^2 = P_2(u_2),
\een
 with
\ben
 P_2(u_2) = A^2 \lambda ^2  P_1\left( \frac{u_2}{A} -B\right) + 2 \lambda ^2 AB u_2 +
(1-\lambda ^2) u_2^2 - \lambda ^2 A^2  B^2.
\een

We may compose two such tranformations to get a third:
\ben
A_3 =  A_2A_1,\qquad \lambda_3 =\lambda _2 \lambda _1,\qquad B_3=B_2 + A_s B_1.
\een
Thus the set of central potentials is acted upon by
a three-parameter group.
The first two formulas are commutative but the second is not.
In the above the second transformation $A_2,B_2,\lambda_2$
follows the first  $A_1,B_1,\lambda_1$.

A matrix representation may be given
\ben
\begin{pmatrix}  u \\ \phi \\  1 \\  \end{pmatrix}   \rightarrow
\begin{pmatrix}  A & 0 & B \\ 0 &\lambda & 0 \\\ 0 &0 & 1 \\ \end{pmatrix}
\begin{pmatrix}  u \\ \phi \\ 1\\ \end{pmatrix}
\een
which reveals the three-dimensional group as  ${\Bbb R} ^\times$
corresponding to $\lambda$ times the semi-direct product
${\Bbb R} ^\times \ltimes   {\Bbb R} ^+$ corresponding to $A$ and $B$.
Acting on $P(u)$, we have
\ben
 P(u) \rightarrow  \lambda ^2  A^2  \left( P\left(\frac{u}{A}  +B \right) - \left(\frac{u}{A}  +B\right)^2  \right )  +u^2.  \een
 Or if $Q(u)=P(u)-u^2$, then
\ben
Q(u) \rightarrow  \lambda ^2  A^2 Q\left(\frac{u}{A}  +B \right),
\een
which says that $Q(u)$  transforms under pull-back
composed with mutiplication by $A^2 \lambda^2$.

\subsection{Non-linear transformations}

The simplest of these is to raise $u$ to a   power: If
\ben
u_2=(u_1)^n,
\een
then
\ben
P_2(u)= n^2 u^{2 \frac{n-1}{n} }  P_1 \left(u^ {\frac{1}{n}}\right) + (1-n^2 )u^2.
\een

The case $n=1$ is particularly simple:
\ben
P_2 (u)= u^4 P_1\left(\frac{1}{u}\right).
\een

\subsection{Chebyshev's theorem}

From (\ref{8pedal}) we deduce that
\ben
\phi= \int \frac{\dd u}{\sqrt{P(u) -u^2}} \,.
\een
Chebyshev's theorem states that  for rational numbers
$p,q,r$ ($r\neq0$) \footnote{$p$ in this section
should not be confused for the perpendicular  distance
from the origin to the tangent.}
and nonzero real numbers $\alpha,\beta$, the integral
\ben
I=\int x^p\left(\alpha+\beta x^r\right)^q\, \dd x = \int
\left(\alpha  x^{\frac{p}{q}} +\beta x^{\frac{p}{q} +r   } \right)^q\, \dd x
\een
 is elementary
if and only if at least one of the quantities
\ben
\frac{p+1}r,\quad q,\quad \frac{p+1}r+q,
\een
is an integer. In our case $q=-\half$ and so integrability in finite terms is possible
if \ben
P(u)= u^2 + \alpha  u^{\frac{p}{q}} +\beta u^{\frac{p}{q} +r   }  \,,
\een

where $\frac{p+1}{r}$ is an integer or half integer.

\subsection{Examples}

\subsubsection{Circles}

With respect to a point distance $A$ from a circle of
radius $R$
\ben
r^2 =A^2 -R^2 + 2 Rp,\qquad \frac{1}{p^2}= \frac{4 R^2 u^4 }
{\left( 1+ (R^2-A^2) u^2  \right )^2}.
\een

\subsubsection{Conics}
An ellipse
\ben
\frac{x^2 }{A^2}+ \frac{y^2}{B^2 } =1,
\een
with respect to the focus
\ben
\frac{1}{p^2} = \frac{1}{B^2} \left ( \frac{1}{2Au}-1 \right),
\een
and with respect to the centre
\ben
\frac{1}{p^2}= \frac{A^2+B^2} {A^2B^2}  -\frac{1}{u^2}.
\een
A hyperbola
\ben
\frac{x^2 }{A^2}-\frac{y^2}{B^2 } =1,
\een
with respect to the focus
\ben
\frac{1}{p^2} = \frac{1}{B^2} \left ( 1- \frac{1}{2Au} \right ),
\een
and with respect to the centre
\ben
\frac{1}{p^2}= \frac{A^2-B^2} {A^2B^2}  + \frac{1}{u^2}.
\een
A parabola
\ben
y^2=4Ax,
\een
with respect to the focus
\ben\frac{1}{p^2}= \frac{u}{A}, \een
with rsepect to its vertex
\ben
A^2(r^2-p^2) ^2 = p^2 (r^2 + 4A^2) (p^2 + 4A ^2 ).
\een

\subsubsection{Algebraic spirals}

\ben
u=\phi ^{-m}, \quad \frac{1}{p^2}= u^2 + \frac{1}{m^2} u^{2m+2}.
\een

\begin{itemize}
\item{$m=1$: Archimedes Spiral.}
\item{$m=1$: Hyperbolic or Reciprocal Spiral.}
\item{$m=2$: Galilei's Spiral.}
\item{$m=-\half$: Cotes's Lituus.}
\item{$m=\half$: Fermat's Spiral.}
\end{itemize}

\subsubsection{Logarithmic or equiangular spiral}

\begin{itemize}
\item
$u= \e^{-\phi \cot\alpha} \Longleftrightarrow  p=r \sin \alpha$,
where $\alpha $ is the angle beween the curve and the radius vector.
\end{itemize}

\subsubsection{Sinusoidal spirals}

\ben u=\left(\cos k \phi \right )^m,\quad
\frac{1}{p^2}=k^2 m^2 u^{ 2(1-\frac{1}{m})}
+ (1-k^2m^2 ) u^2.
\een

\noindent If $m=-1$ {we obtain the  Rhodenea} $u= \frac{1}{\cos k \phi} $.

\medskip
\noindent $k=3$ gives the  {Trifolium}.

\medskip
\noindent  $k=2$ {the  Quadrifolium} and

\medskip \noindent $k=\frac{1}{3}$ is the {pedal of the Cardioid}.

\medskip
\noindent If $m=1$ we obtain the {Epi-spirals} $u= {\cos k \phi} $.

\medskip
\noindent \medskip If $m=-\frac{1}{k}$ we have $u= (\cos k \phi )^{-\frac{1}{k}}$
and $\frac{1}{p^2} = u^{2(1+k)} $.

Some particlular sinusoidal spirals  are:
\begin{itemize}

\item $m=1,k=\half$.  {The  Trisectrix of De Longe}
$u=\cos \frac{\phi}{2}$,     $\frac{1}{p^2} = \frac{1}{4} + \frac{3}{4} u^2 $.

\item   $m=1, k= \frac{1}{3} $.  { Maclaurin's Trisectrix}
 $u=\cos \frac{\phi}{3} $,
 $ \frac{1}{p^2} = \frac{1}{9} + \frac{8}{9} u^2  $.

\item $m=3, k=1 $.  {The  Cubic Duplicatrix} $u=\cos ^3 \phi $,
 $ \frac{1}{p^2} = 9 u^{\frac{4}{3}} -8 u^2   $.

\item  $m=2, k=1 $.     {The  Kampyle }
 $u=\cos ^2 \phi $,
  $ \frac{1}{p^2} = 4 u  -3 u^2   $.

\item  $m= \frac{3}{2}, k=1 $.  {The  Witch of Agnesi  }
 $u=\cos ^{\frac{3}{2}}  \phi $,
 $ \frac{1}{p^2} = 4 u ^{\frac{2}{3} } - \frac{5}{4} u^2$.

\item $m= 2, k=1 $. {The Cruciform} $u=\cos 2  \phi $,
 $ \frac{1}{p^2} = 4  - 3 u^2  $.

\item $m= -2, k=1 $. {The Oeuf Double } $u=\cos ^{-2}    \phi $,
$ \frac{1}{p^2} = 4 u^3      - 3 u^2  $.

\item $m=-3, k=\frac{1}{3}$. { Cayley's Sextic} $u=\frac{1}{\cos ^3\frac{\phi}{3}} $, $ \frac{1}{p^2} = u^{\frac{8}{3}}.$

\end{itemize}

\subsubsection{Black holes}

As central orbits, the  first  two examples below are related by
Arnold--Bohlin duality \cite{Arnold,Bohlin} while the third is self-dual.
They arise  in the theory of null geodesics of
spherically symmetric Ricci flat black hole metrics \cite{Gibbons:2011rh}.
\medskip

\noindent If $u= \frac{\cosh \phi +2}{\cosh \phi-1}$ or
$u= \frac{\cosh \phi -2}{\cosh \phi+1}$
we have  $\frac{1}{p^2} = \frac{1}{3} + \frac{2}{3} u^3 $.
These are particular null geodesics of a four-dimensional
black hole.

\medskip
\noindent If $u^2 = \frac{\cosh2 \phi +1}{\cosh 2\phi-2}$ or
$u= \frac{\cosh 2\phi -1}{\cosh \phi+2}$
we have  $\frac{1}{p^2} = \frac{2}{3} + \frac{1}{3} u^6 $.
These are particular null geodesics of a seven-dimensional
black hole.

\medskip
\noindent If $u = \tanh \left( \frac{\phi}{\sqrt{2}}\right)$ or
$u = \coth \left( \frac{\phi}{\sqrt{2}}\right)   $
we have  $\frac{1}{p^2} = 2(1+ u^4)$.
These are particular null geodesics of a five-dimensional
black hole.

\subsubsection{Other examples}
Freeth's Nephroid is
\ben
r= \left(1+ 2 \sin \frac{\phi}{2} \right),
\een
whence
\ben
\frac{1}{p^2}= \frac{u^2}{4} ( 7 -2u + u^2 ).
\een

\subsubsection{Some hyperbolic spirals }

\ben
u=\left( \tanh k \phi \right )^m,
u=\left( \coth k \phi \right )^m, \frac{1}{p^2} =
m^2k^2 \left( u^{2 \frac{m-1}{m}}   + u^{2\frac{m+1}{m} }  \right )
+ (1-2 m^2 k^2)  u^2.
\een

\ben
u= \left(\cosh k\phi \right )^m,
  \quad \frac{1}{p^2} = -m^2k^2 u^{2\frac{m-1}{m} } + (1+m^2k^2) u^2,
\een
\ben
u= \left(\sinh k\phi \right )^m,
  \quad \frac{1}{p^2} = m^2k^2 u^{2\frac{m-1}{m} } + (1+m^2k^2) u^2.
\een
If $m=1$ we get the two types of Poinsot Spirals
for which
\ben
\frac{1}{p^2}= (k^2 +1)^2 \mp k^2,
\een
respectivley.

\subsubsection{Further examples}

\begin{itemize}

\item The astroid is
\ben
\frac{1}{p^2}= \frac{3 u^2}{u^2-1}.
\een 
\item

The nehproid is
\ben
\frac{1}{p^2}= \frac{3 u^2}{4-16u^2}.
\een

\item The deltoid is
\ben
\frac{1}{p^2}= \frac{8 u^2}{9u^2-1}.
\een

\end{itemize}

\ben
u= \left( \tan k \phi \right )^m,\quad \frac{1}{p^2}=
m^2 k^2 \left(  u^{2(1+\frac{1}{m}) } +   u^{2 (1-\frac{1}{m}) } \right )  +
(2 m^2 k^2 +1 )  u^2.
\een

\begin{itemize}

\item { Kappa } $u=\cot\phi $, $m=-1, k=1$,
   $ \frac{1}{p^2} = 1+ 3 u^2 + u^4 $.

\item { Moulin \'a vent  } $u= \cot2 \phi  $,
$m= 1, k=2 $,
 $ \frac{1}{p^2} =4+  9u^2  + 4 u^4     $.

\end{itemize}

\end{document}